\documentclass[conference]{IEEEtran}
\IEEEoverridecommandlockouts
\usepackage{cite}
\usepackage{amsmath, amssymb, amsfonts, mathrsfs}
\usepackage{algorithmic}
\usepackage{times}
\usepackage[margin=1in]{geometry}
\usepackage[normalem]{ulem} 
\usepackage{graphicx, color}
\usepackage{textcomp}
\usepackage{xcolor}
\usepackage{multirow}

\newcommand{\eat}[1]{}

\begin{document}

\title{Integrating Machine Learning with HPC-driven Simulations for Enhanced Student Learning\\}

\author{\IEEEauthorblockN{
Vikram Jadhao,
JCS Kadupitiya}
\IEEEauthorblockA{\textit{Intelligent Systems Engineering} \\
\textit{Indiana University}\\
Bloomington, Indiana 47408 \\
\{vjadhao,kadu\}@iu.edu}
}

\maketitle

\begin{abstract}
We explore the idea of integrating machine learning (ML) with high performance computing (HPC)-driven simulations to address challenges in using simulations to teach computational science and engineering courses. We demonstrate that a ML surrogate, designed using artificial neural networks, yields predictions in excellent agreement with explicit simulation, but at far less time and computing costs. 
We develop a web application on nanoHUB that supports both HPC-driven simulation and the ML surrogate methods to produce simulation outputs. 
This tool is used for both in-classroom instruction and for solving homework problems associated with two courses covering topics in the broad areas of computational materials science, modeling and simulation, and engineering applications of HPC-enabled simulations.
The evaluation of the tool via in-classroom student feedback and surveys shows that the ML-enhanced tool provides a dynamic and responsive simulation environment that enhances student learning. 
The improvement in the interactivity with the simulation framework in terms of real-time engagement and anytime access enables students to develop intuition for the physical system behavior through rapid visualization of variations in output quantities with changes in inputs. 
\end{abstract}


\section{Introduction}\label{sec.intro}
The use of computational simulations is ubiquitous in investigating phenomena associated with a wide range of disciplines including materials science and engineering, bioengineering, chemistry, chemical engineering, and physics. Simulations have enabled the understanding of microscopic mechanisms underlying several biological and material phenomena such as ion transport across the cell membrane, flow of polymeric liquids, stabilization of colloidal dispersions, and self-assembly of nanostructures \cite{frenkel,glotzer2015assembly,brunk2019linker,jing2015ionic,jadhao2019rheological}. 
Classical molecular dynamics (MD) simulations are an important class of simulation approaches that are generalizable to study a broad range of material and chemical systems \cite{frenkel}. In the MD method, Newton's equations of motion for a system of many particles are solved at each timestep to evolve the particle positions, velocities, and forces forward in time. In several applications where the computational complexity per time step is proportional to the square of the total number of particles (system size), MD simulations incur high computational costs. These high costs are typically mitigated by employing high performance computing (HPC) resources and utilizing parallel computing techniques such as OpenMP and MPI. Using HPC-enabled acceleration techniques can dramatically enhance the performance of MD simulations in many cases.

The parallelized MD simulations enable dynamics of systems with a large number of particles over a wide range of input system parameters. In addition to enabling state-of-the-art research, these simulations can be employed as innovative educational tools for teaching materials covered in computational science and engineering courses. However, despite the employment of the optimal parallelization model suited for the size and complexity of the model system, MD simulations can take a relatively long time to furnish accurate information, varying from minutes to days depending on the model system specifics. Primary factors contributing to this scenario are the time delays resulting from the combination of waiting time in a queue on a computing cluster and the actual runtime for the simulation. 
Given this prognosis, the use of MD simulations has been generally limited to ``outside-classroom'' activities such as solving homework problems, where the simulation time requirements are easier to meet.

To use MD simulations during in-classroom sessions, the associated computational tool needs to be sufficiently agile to overcome the following educational challenges:
\begin{itemize}
    \item Provide simulation-based responses to student questions \emph{in real-time}.
    \item Make the process of explaining underlying scientific concepts seamless by having \emph{rapid access to accurate trends} in the variation of simulation outputs. 
    \item Do \emph{synchronous} simulation-based analysis of model system behavior in real-time with students.   
    \item Provide a \emph{dynamic environment} for students to perform \emph{in silico} experiments during class to learn the concepts by visualizing the system response under different input conditions. 
\end{itemize}
Addressing these educational challenges can improve the classroom teaching of computational science and engineering concepts, and enhance student learning.

Motivated in part by these challenges, we recently introduced the idea of integrating machine learning (ML) methods with MD simulations to develop ``ML surrogates'' for MD simulations \cite{kadupitiya2019machine,kadupitiya2020machine2}. 
We demonstrated that an artificial neural network (ANN) based regression model, trained on data produced by completed runs of a given HPC-accelerated MD simulation, can successfully imitate part or all of that MD simulation. We also showed that performance improvements of several orders of magnitude could be achieved by replacing traditional large-scale HPC simulations with ML surrogates. The central idea and approach were illustrated using MD simulations of ions in nanoconfinement \cite{jing2015ionic}. The ML surrogate was found to accurately predict the ionic distributions in confinement and produce the outputs with an inference time of over a factor $10,000$ smaller than the corresponding MD simulation runtime \cite{kadupitige2019machine,kadupitiya2020machine2}.

Our earlier papers focused on the technical details of designing ML surrogates for MD simulations and evaluating the performance metrics associated with the proposed data-driven approach \cite{kadupitige2019machine,kadupitiya2020machine2}. In this paper, we explore the potential of employing these ML surrogates to address the aforementioned educational challenges and enhance the usability of simulation tools in education. We develop a nanoHUB web application with a GUI that supports both ML surrogate and MD simulation methods to produce simulation outputs. 
The nanoHUB tool is used for both in-classroom teaching and for solving homework problems associated with two courses offered at Indiana University (IU). The ML surrogate built into the tool is employed extensively during the in-classroom instruction to teach concepts such as self-assembly, ionic behavior near interfaces, nanoscale material design, modeling and simulation, and neural networks. 
The impact of the use of ML-enhanced tool in student learning is assessed and evaluated by conducting a survey following other assessment studies \cite{tanaka2019teaching,srivastava2019assessing}.
Based on the educational evaluation of the tool, we find that the improvement in the interactivity with the simulation framework in terms of dynamic, real-time engagement and anytime access enables enhanced student learning of computational science concepts.

\section{Related Work}

\subsection{ML for enhancing simulations of material systems}
Computational science and engineering is being transformed by the use of ML. In the area of simulations of materials, ML techniques and in particular deep neural networks based methods have been used to predict parameters, generate configurations, classify material properties, and design force fields \cite{glotzer2017,sam2017,ferguson2017machine,melko2017, kadupitiya2018machine,fox2019learning,wang2019machine}. More recently, ML has been used to design surrogate models that can predict specific outcomes of simulations by bypassing part or all of the simulation. For example, the dissociation timescale of compounds was predicted using an ML surrogate for \emph{ab initio} MD simulations by bypassing the time evolution of the particle trajectories \cite{aspuru2019}. Deep neural networks trained on HPC-generated simulation data were used as an efficient surrogate for molecular simulation to predict adsorption equilibria as a function of thermodynamic state variables \cite{sun2019deep}. Convolutional neural network based ML ``emulators'' have been developed to predict simulation outputs such as power spectrum in biogeochemistry \cite{kasim2020up}. We have also developed surrogates that can predict the outputs (ionic density profiles) of MD simulations of ions in confinement \cite{kadupitiya2019machine,kadupitiya2020machine2}, or in another example, compute forces at each timestep in an MD simulation to bypass the expensive force calculation step \cite{kadupitiya2020simulating}.
While these ML surrogates have enabled remarkable performance improvements to facilitate research investigations, their use for education applications has been relatively unexplored despite their ability to produce outputs in real-time and for a continuous range of input parameters. In this paper, we probe the potential of using ML surrogates for MD simulations to enhance student learning of topics in the area of computational science and engineering.

\subsection{Simulation caching}
Generally, the approach to provide simulation output in real-time is to store the previous simulation results in a cache (simulation caching). For example, nanoHUB provides caching as a feature in computational tools created using their Rappture GUI \cite{klimeck.nanohub}. Cached simulations provide a static environment with pre-selected parameters defining simulations that can be ``looked up''. This simulation environment offers limited exploration space, interactivity, and responsiveness to the student. To encourage and empower students to directly experiment and explore the model system and associated phenomena, a new approach is needed that delivers an interactive, dynamic, and responsive simulation environment open for wide exploration. We show that ML surrogates are excellent candidates to fulfill this need.

\section{Background}

\subsection{Use of HPC-enabled simulations in education}

One of us teaches two courses at IU that have been taken by undergraduate and graduate students with interests in diverse focus areas including nanoscale engineering, bioengineering, computer engineering, chemistry, and physics. The courses feature application-based learning of basic scientific computing concepts and simulation techniques, including the use of parallel computing methods. Applications are designed employing the state-of-the-art research in nanomaterials engineering covering several material systems such as virus-like particles \cite{brunk2019linker}, shape-changing nanocontainers \cite{brunk2019computational,jadhao2015coulomb}, ion channels \cite{jing2015ionic}, and polymeric fluids \cite{jadhao2019rheological}. HPC-enabled MD simulations are key parts of these courses. These  simulations serve as important tools for understanding diverse self-assembly phenomena in nanoscale materials \cite{glotzer2015assembly}, predicting material behavior in practical applications \cite{jadhao2019rheological}, and isolating interesting regions of parameter space for experimental exploration \cite{brunk2019computational}. 

\subsection{GUI-wrapped simulations on nanoHUB}

To facilitate the use of simulations by students in classrooms and for solving homework problems, the simulations are deployed as computational tools on nanoHUB \cite{klimeck.nanohub}. nanoHUB provides free online access and a Jupyter-notebook based GUI wrapper for executing simulation codes. Depending on the input selected by the users on the GUI, simulations are launched on virtual machines or on a supercomputing cluster such that the associated simulation wait and run time is minimized. The authors and the extended research group members have published 6 nanoHUB tools that enable exploration of diverse self-assembly phenomena in nanomaterials: Ions in nanoconfinement \cite{kadupitiya2017}, Nanosphere electrostatics lab \cite{kadupitiya2018}, Nanoparticle assembly lab \cite{brunk2019}, Nanoparticle shape lab \cite{brunk2020}, Polyvalent nanoparticle binding simulator \cite{lauren2019}, and Souffle: Virus capsid assembly lab \cite{lauren2020}.  
These tools provide an interactive GUI to students for examining the links between nanomaterial system parameters and their structural and dynamical behavior via renderings of simulation output on the tool canvas. The tools also enable students to learn the workflows associated with a large scientific simulation software ecosystem. Three of the six tools have already been employed in teaching materials associated with the aforementioned courses. Some in-classroom lecture videos are available on nanoHUB as educational resources \cite{video.nano,video.shape}.

\subsection{nanoHUB tool: Ions in nanoconfinement}

The nanoHUB tool ``Ions in nanoconfinement'' \cite{kadupitiya2017} enables users to simulate the self-assembly of ions in nanoconfinement created by material surfaces represented as identical planar interfaces. 
A primitive model of electrolytes is used to model the ions \cite{jing2015ionic,solis2013generating}. Simulations are performed using standard molecular dynamics methods \cite{jing2015ionic,lammps.plimpton}.
The inputs associated with the tool include ion valency, ion size, electrolyte concentration, and interface separation. The simulation outputs include the density profiles of ions in confinement. 
This tool has been employed every semester since Spring 2018 in illustrating concepts in a graduate course (Simulating Nanoscale Systems) and an undergraduate course (Introduction to Modeling and Simulation) at IU. In less than 3 years of its launch, the tool has been used by over 130 users and run over 3400 times \cite{kadupitiya2017}. 

\begin{figure}[ht]
\centerline{\includegraphics[scale=0.125]{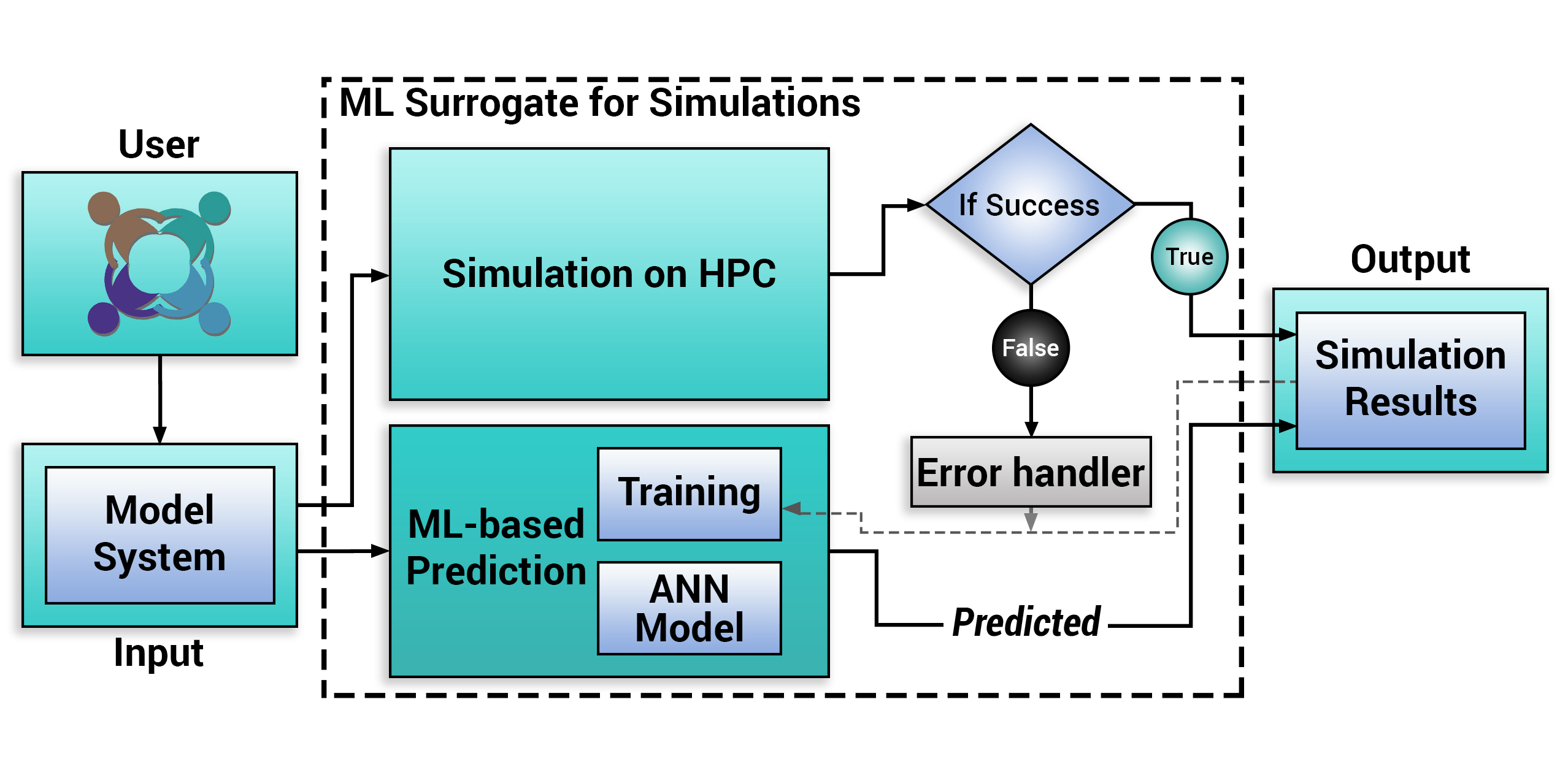}}
\caption
{\label{fig.overview}
System overview of the ML surrogate for simulation approach for generating rapid and accurate predictions of simulation outputs for use in classroom teaching.
}
\end{figure}

The tool comes with a hybrid OpenMP/MPI acceleration that enables simulations to be completed within 30 minutes for any combination of input parameters (assuming no waiting time on HPC cluster). In classroom usage, we observed that the fastest simulations took about 10 minutes to provide the converged ionic densities while the slowest ones (generally associated with a larger number of simulation steps and system sizes) took as long as 1 hour. 
Primary factors contributing to this scenario were the time delays resulting from the combination of waiting time in a queue on a computing cluster and the actual runtime of the MD simulation. Not having rapid access to expected trends in the variation of ionic densities with input parameters made the process of explaining concepts and mechanisms unwieldy and time-consuming. 
As we demonstrate below, integrating this tool with a ML surrogate improved its overall usability as an educational tool.

\begin{figure*}[htb]
\centerline{\includegraphics[scale=0.27]{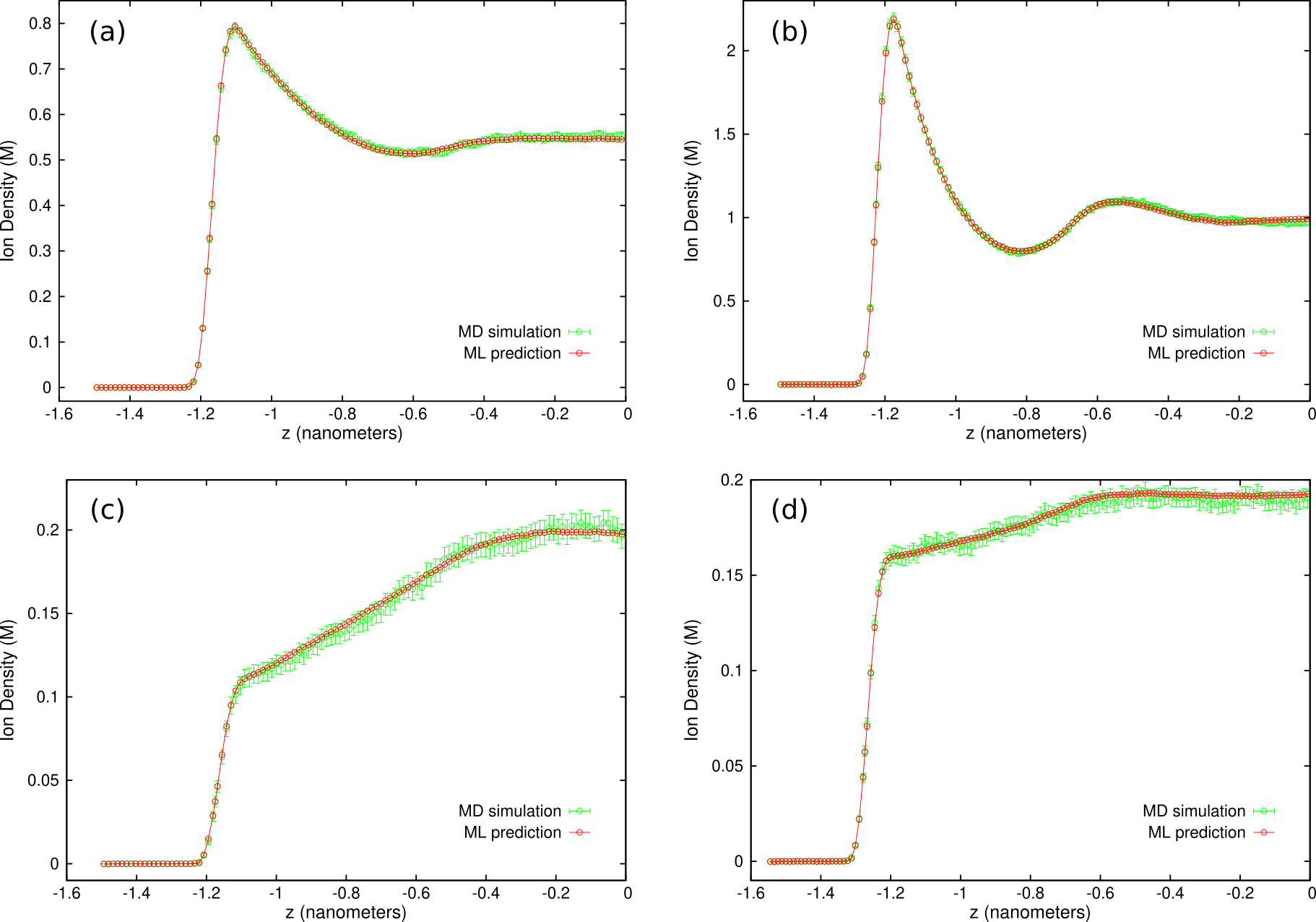}}
\caption
{\label{fig.predictions}
Ionic density profiles for systems I (a), II (b), III (c), and IV (d) predicted by the ML surrogate (red circles) and extracted with MD simulation (green circles with errorbars). See main text for system definitions. For each system, the ML-predicted density profile is in excellent agreement with the simulation result.
}
\end{figure*}

\section{ML Surrogates for Simulations}

We now describe a general approach, introduced in Refs. \cite{kadupitiya2019machine,kadupitiya2020machine2}, that utilizes ML to enable real-time and anytime engagement with simulations, significantly enhancing the potential for their use in both research and education. In this approach, ML surrogate model is designed using data produced by completed runs of a given HPC-accelerated simulation, and then deployed to approximate the complex relationships between the physical input parameters and the output results of simulations. The ML surrogate bypasses the explicit computational evolution of the simulated model components, yielding accurate outputs in much less time and computing costs. Figure \ref{fig.overview} shows the overview of this framework. First, the attributes of the model system are fed to the framework. These inputs are used to launch the simulation on the HPC cluster. Simultaneously, these inputs are fed to the ML-based prediction module. Both the simulation and ML methods are designed to extract (infer) the desired output quantities. Error handler aborts the simulation program and displays appropriate error messages when a simulation fails due to any pre-defined criteria. At the end of the simulation run, the output quantities are saved for future retraining of the ML model, which occurs after a set number of new successful simulation runs. 

In previous papers, we applied this framework to the case of MD simulations of ions in nanoconfinement.
The surrogate was trained to learn the relationship between the output distribution of positive ions and 5 input parameters characterizing the ionic system: confinement length $h$, salt concentration $c$, positive ion valency $z_p$, negative ion valency $z_n$, and ion diameter $d$.
The range of each input parameter were as follows: $h \in (3.0, 4.0)$ nm, $c \in (0.3, 0.9)$ M, $z_p \in{1,2,3}$, $z_n \in{-1}$, and $d \in (0.5, 0.75)$ nm. The output quantity was selected to be the distribution of positive ions confined by two identical planar interfaces at $z = -h/2$ and $z = h/2$. For simplicity, using the symmetry of the ionic density around the confinement center $z = 0$, ML surrogate was trained to make predictions characterizing the density of ions in the left half of the confinement (i.e., for $z \in (-h/2, 0)$). The predictions were made at approximately $150$ positions; the associated $P \approx 150$ density values were selected as the output parameters (features). 

The dataset for training the ANN-based ML surrogate was generated by sweeping over a few discrete values for each of the input and output parameters to create and run 6,864 MD simulations utilizing HPC resources. On average, each MD simulation was run for over $5$ million computational steps and took 4200 CPU hours ($\approx 36$ minutes per simulation). The training dataset creation took approximately 25 days, including the queue wait times on the IU BigRed2 supercomputing cluster. The entire data set was separated into training and testing sets using a ratio of 0.8:0.2. The ANN architecture with 2 hidden layers was implemented in Python using scikit-learn, Keras, and TensorFlow ML libraries \cite{chollet2015keras,buitinck2013api,abadi2016tensorflow} for regression and prediction of $P \approx 150$ continuous (output) variables. The details of the data generation, preprocessing, ANN feature extraction, and regression are provided in our earlier papers \cite{kadupitige2019machine,kadupitiya2020machine2}.

The ANN-based surrogate model produced the ionic distribution in excellent agreement with explicit MD simulation results \cite{kadupitiya2020machine2}. In addition to the high accuracy of ML inferences, the surrogate yielded output results over 10,000 times faster than the parallel MD simulation. The typical ML inference time associated with a prediction of the density profile was $\approx 0.2$ seconds (or, almost instantaneous). In strike contrast, the average runtime of the parallel MD simulation to produce a similarly converged output was $\approx 36$ minutes.

\begin{figure*}[ht]
\centerline{\includegraphics[scale=0.335]{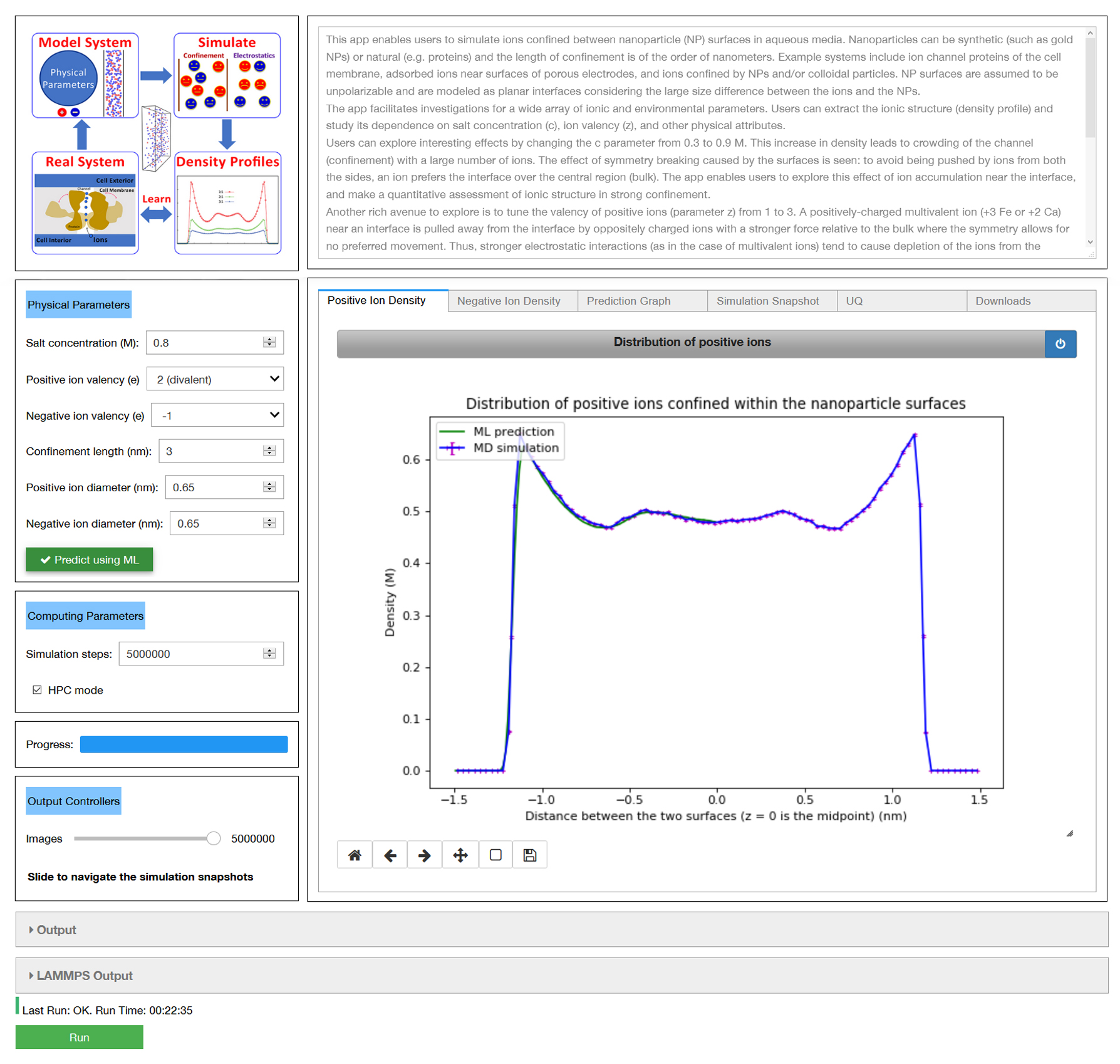}}
\caption
{\label{fig.app_screen_cap}
GUI of the ML-enhanced ``Ions in nanoconfinement'' nanoHUB tool \cite{kadupitiya2017}. The GUI shows the density profile predicted by the ML surrogate for half of the position values (green line) and the result extracted via MD simulation (red markers) for an example ionic system. 
}
\end{figure*}

The overall success rates and rapid inference times associated with predictions made by ML surrogates enable a dynamic and responsive simulation environment for exploration by students in classroom settings. The following capabilities associated with these surrogates are of particular significance in education use: 
\begin{itemize}
\item Learning pre-identified features associated with the simulation outputs 
\item Generating accurate predictions in real-time for unsimulated state points
\item Providing a dynamic environment to rapidly explore the input-output relationships
\item Enabling anytime and anywhere access to simulation results
\end{itemize}
In the next section, we describe the results associated with the use of simulation tools integrated with ML surrogates in teaching materials associated with computational science and engineering courses.

\section{Results}

\subsection{Technical evaluation}

We first discuss  the technical results showing the comparison between the predictions made by the ML surrogate and the outputs of MD simulations. Figure \ref{fig.predictions} (a) - (d) shows the ionic density profiles predicted by the ML surrogate for a set of 4 systems randomly selected from the entire testing dataset. These systems are: system I $(3, 1, -1, 0.45, 0.7)$, system II $(3.6, 1, -1, 0.9, 0.714)$, system III $(3.7, 3, -1, 0.45, 0.5)$, and system IV $(3.3, 2, -1, 0.3, 0.553)$, where the parentheses list the 5 aforementioned input parameters characterizing the ionic system: confinement length $h$, positive ion valency $z_p$, negative ion valency $z_n$, salt concentration $c$, and ion diameter $d$. 
As the figure indicates, for each system, the ML-predicted density profile is in excellent agreement with the result extracted using MD simulation (ground truth).
In addition to the high accuracy, we note that the ML inferences are made in a much shorter time of $\approx 0.2$ seconds compared to MD simulations ($\approx 36$ minutes on average).

Motivated by the good agreement between ionic densities generated via ML surrogate and MD simulations as well as the remarkable performance enhancement resulting from the use of ML surrogates, we integrated the ML surrogate with the nanoHUB tool ``Ions in nanoconfinement''. The ML-enhanced tool was deployed on nanoHUB in October 2019. 
Figure~\ref{fig.app_screen_cap} shows the Jupyter python notebook based GUI of the deployed tool. Users are provided with the choice to click ``Run'' and ``Predict using ML'' buttons simultaneously or separately depending on the desired information. ``Predict using ML'' activates the ML surrogate which predicts half of the density profile instantaneously; the result is available in the ``Prediction Graph'' tab as well as in the ``Positive Ion Density'' tab. Users can enable ML surrogate any time by clicking the ``Predict using ML'' button to access the ML-predicted ionic density profile.
Clicking the ``Run'' button instructs the execution engine to either submit a job on an HPC cluster (if the ``Cluster mode'' button is checked) or run the simulation on a VM. When the simulation is over, the execution engine passes the generated data to be plotted on the ``Positive Ion Density'' and ``Negative Ion Density'' tabs. 
For illustration purposes, Figure~\ref{fig.app_screen_cap} also shows the final density plot obtained using the integrated MD and ML method for the input parameters $h = 3.0$ nm,  $z_p = 1$,  $z_n = -1$, $c = 0.5$ M, and $d = 0.714$ nm. The ML prediction is shown as an overlay in the ``Positive Ion Density" tab along with the result of the MD simulation. 

\subsection{Educational evaluation}

\begin{figure}[htb]
\centering
\includegraphics[scale=0.64]{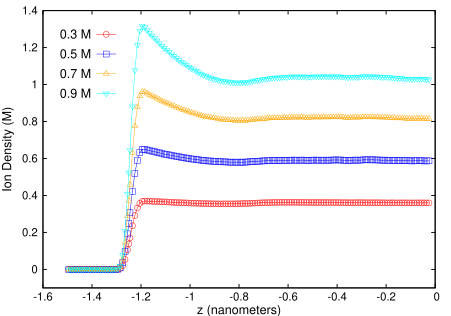}
\caption{Density profiles of confined positive ions for different salt concentration $c = 0.3, 0.5, 0.7, 0.9$ M predicted by the ML surrogate.}
\label{fig.t3}
\end{figure}

An accurate and rapid assessment of ionic distributions in confinement by the ML surrogate enables in-classroom instruction of several important concepts such as interfacial effects, self-assembly in nanoscale systems, and the intimate connection between solution conditions and the material assembly behavior. 
For example, by using the ML surrogate, students can instantaneously record changes in the ionic structure as the salt concentration $c$ is tuned. 
Figure \ref{fig.t3} shows a selected subset of ionic density profiles predicted by the ML surrogate for different $c = 0.3, 0.5, 0.7, 0.9$ M. 
Other input parameters are fixed to $h = 3.0$ nm, $z_p = 1$, $z_n = -1$, and $d = 0.5$ nm. 
By performing \emph{in silico} experiments in rapid succession using the ML surrogate, students can readily visualize the response of the ionic system under changes in $c$. For example, students learn that increasing salt concentration leads to the accumulation of ions near the interface (higher peaks in the ionic density) or to the emergence of more modulations in the density profile.
Both these observations inferred by the ML surrogate follow the expected behavior in these systems as reported and elucidated in previous work \cite{jing2015ionic}. 
The ML surrogate also enables instructors to perform simulation-based analysis of the ionic model system behavior synchronously with students. Further, the remarkable agility of the surrogate in yielding the predictions enables educators to respond to student questions in real-time via live demonstration using the ML surrogate.
The ML surrogate thus helps resolve the 4 key educational challenges outlined in the Introduction (Section \ref{sec.intro}).

As noted before, one of the authors regularly teaches two courses at IU: 1) Simulating nanoscale systems (Fall semester) and 2) Introduction to modeling and simulation (Spring semester). 
The students in these courses learn computational model development, simulation techniques such as molecular dynamics, data analysis and visualization, computational materials science concepts such as self-assembly and interfacial phenomena, parallel computing methods, and engineering applications of simulations. The learning is facilitated by having students perform HPC-based simulations that enable the extraction of structure-property relationships in materials at the nanoscale. Students also become familiar with important practical aspects of research in scientific computing such as scalability, time discretization, convergence, model resolution, and simulation accuracy. 

nanoHUB computational tools are key parts of these courses as they help facilitate the use of simulations by students via a user-friendly, web application requiring no software installation to run the simulations.
The ML surrogate was integrated into the nanoHUB tool ``Ions in nanoconfinement'' and the enhanced tool was pilot tested in Course 1 in Fall 2019. Six students majoring in different fields, including nanoengineering, computer engineering, and chemistry took the course. The tool was also used in Course 2 by 15 students in Spring 2020. Students used the tool during in-classroom lectures as well as to solve homework problems. The tool was actively employed by the instructor in the classroom to help students develop an intuitive understanding of the ionic system behavior via rapid experimentation and visualization of changes in ionic structure.

Below we enumerate a subset of the learning outcomes of these courses in order to provide the context for the results of the tool evaluation discussed in the remainder of this section. When students complete the aforementioned two courses they should be able to:
\begin{enumerate}
    \item Develop scale-appropriate and computationally-efficient models of real / experimental systems.
    \item Develop an in-depth understanding of computational materials science concepts such as self-assembly and structure-property relationships.
    \item Develop simulation methods and apply them to solve engineering problems.
    \item Use parallel computing methods to enhance computational simulations of nanoscale materials.
    \item Use web-based computational tools and understand the associated scientific workflow.
\end{enumerate}

\begin{table}[ht]
\caption{Rating-based questions used in the survey}
\begin{center}
\begin{tabular}{|c|c|}
\hline
\textbf{ID} & \textbf{Question} \\
\hline
\hline
\multirow{2}{*}{Q1} & \multirow{2}{0.41\textwidth}{Were the use of simulations in the class valuable in learning concepts?} \\ & \\
\hline
\multirow{2}{*}{Q2} & \multirow{2}{0.41\textwidth}{Were the learning objectives regarding the use of the nanoHUB tool clear?}\\ & \\
\hline
\multirow{2}{*}{Q3} & \multirow{2}{0.41\textwidth}{Rate the tool in terms of user-friendliness.}\\ & \\
\hline
\multirow{2}{*}{Q4} & \multirow{2}{0.41\textwidth}{Rate the tool in terms of convenience (in terms of how fast the results were inferred by ML).}\\ & \\
\hline
\multirow{2}{*}{Q5} & \multirow{2}{0.41\textwidth}{Rate the tool in terms of accuracy (as compared with MD results).}\\ & \\
\hline
\multirow{2}{*}{Q6} & \multirow{2}{0.41\textwidth}{Rate the tool in terms of use for in-class conceptual understanding.}\\ & \\
\hline
\multirow{2}{*}{Q7} & \multirow{2}{0.41\textwidth}{Rate the tool in terms of use for homework problem solving.}\\ & \\
\hline
\multirow{2}{*}{Q8} & \multirow{2}{0.41\textwidth}{Rate the tool in terms of GUI layout.}\\ & \\
\hline
\multirow{2}{*}{Q9} & \multirow{2}{0.41\textwidth}{Rate the tool in terms of quality.}\\ & \\
\hline
\multirow{2}{*}{Q10} & \multirow{2}{0.41\textwidth}{Rate the tool in terms of consistency.}\\ & \\
\hline
\end{tabular}
\end{center}
\label{tab.rating.questions}
\end{table}

To assess the impact of the use of the ML-enhanced tool and associated simulations on student learning, a tool evaluation survey was conducted at the end of the Fall 2019 semester, where students provided feedback on their experience with the tool. The survey questions were constructed following similar educational evaluation studies \cite{tanaka2019teaching, marchant2019teaching, shoker2019successful, srivastava2019assessing, gonzalez2019toward, carratala2019teaching, qasem2019gentle, miller2019measuring}, and comprised of both rating-based as well as text-response-based questions.

\begin{figure}
\centerline{\includegraphics[scale=0.45]{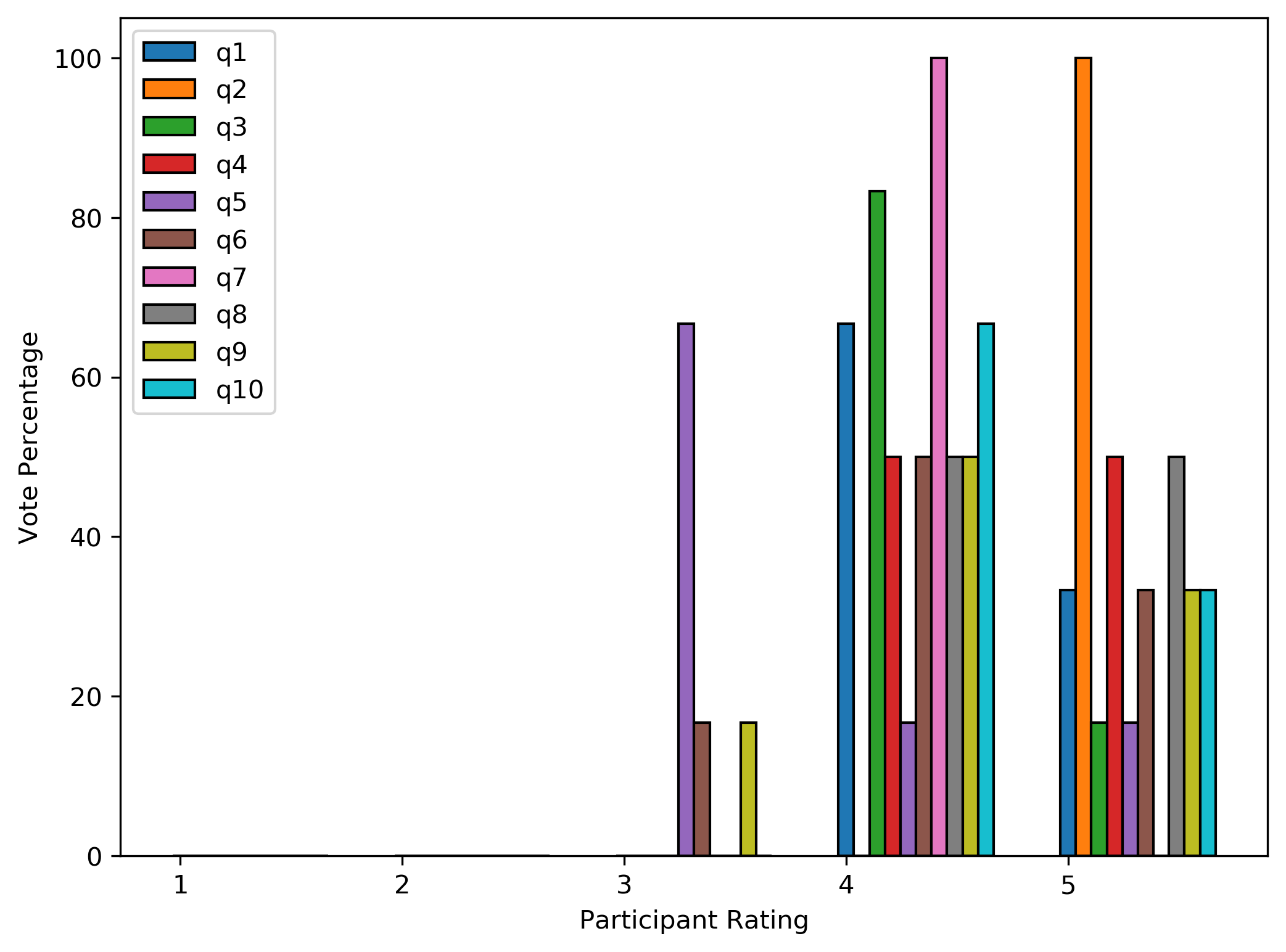}}
\caption
{\label{fig.participant_ratings}
Participant ratings (as a percentage) for the questions (Table \ref{tab.rating.questions}) used in the survey for evaluating the ML-enhanced tool.
}
\end{figure}

\begin{figure}
\centerline{\includegraphics[scale=0.55]{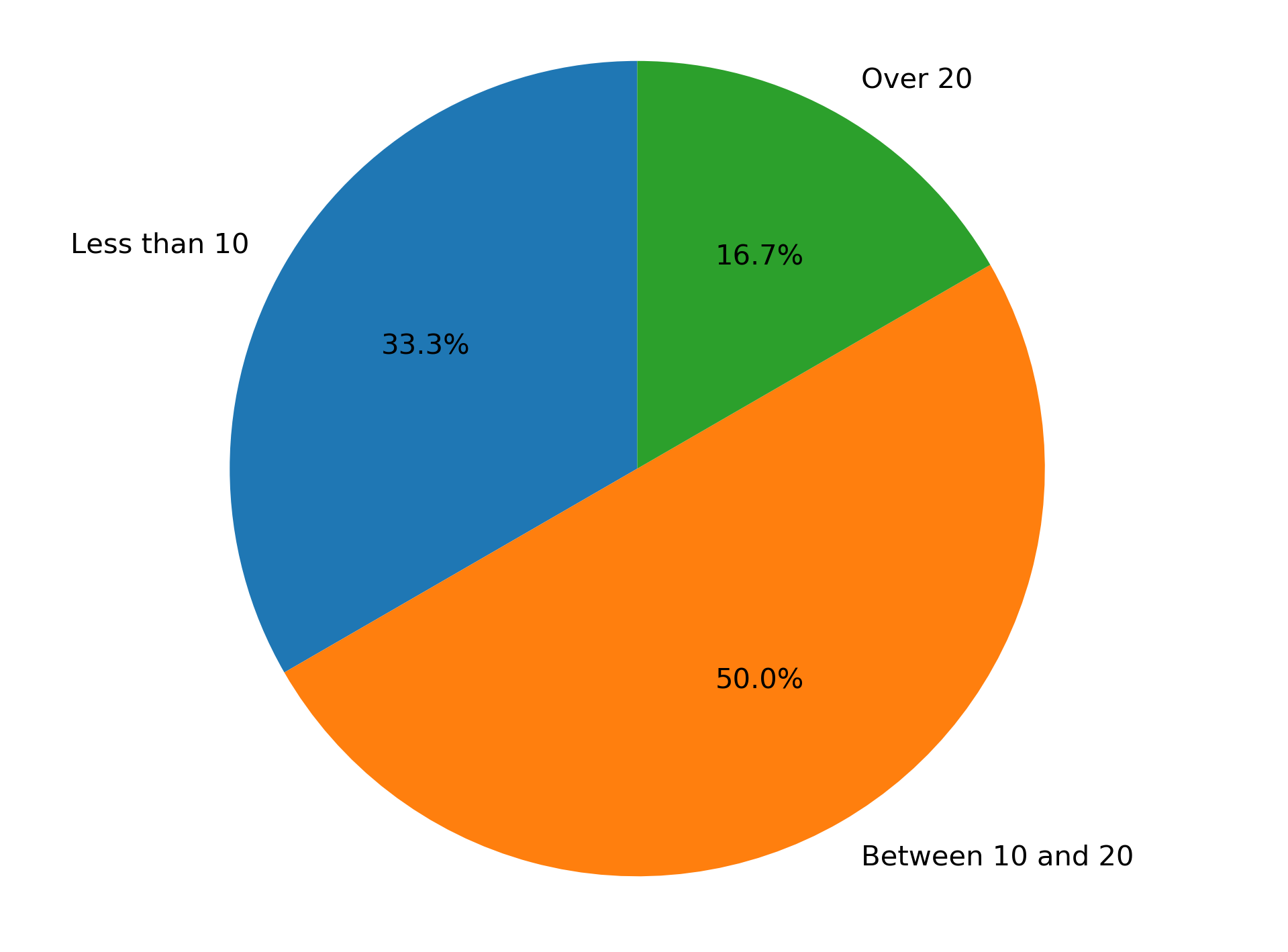}}
\caption
{\label{fig.how_many_times}
A pie chart showing how many times students used the ML-enhanced simulation tool in the classroom. Nearly $16\%$ of the students used the tool over 20 times. 
}
\end{figure}

First, a set of 10 questions tabulated in Table \ref{tab.rating.questions} asked the students to rate the simulation tool in terms of different features such as user-friendliness, clarity, utility, consistency etc.
Participant ratings as a percentage for the 10 questions listed in Table \ref{tab.rating.questions} are shown in the form of a  bar graph in Figure \ref{fig.participant_ratings}. The rating scale was from 1 to 5, where higher scores represent higher ratings. We received a total of 60 rating responses for these 10 questions.
Based on the responses received for all 10 questions, the mean response rating was 4.26 with a variance of 0.39, indicating that on average the students evaluated the simulation tool close to the highest rating of 5.0. 
More specifically, in terms of user-friendliness, convenience, GUI layout, and consistency, students rated the tool at 4 or higher. We also asked the students how many times they used the nanoHUB tool in the class. 16.7\% of the students responded that they have used it more than 20 times, while 50\% said they used the tool between 10 and 20 times (Figure \ref{fig.how_many_times}).

Next, we asked a series of text-response-based questions. We discuss a few of these questions below. 
Students were asked what aspects of the online simulation tool were useful and valuable to them. The students highlighted that the simulation tool helped them increase the conceptual and practical understanding of the nanoscale simulations due to the user-friendly interface, ML-enabled instantaneous predictions, and availability of multidimensional input choices. For example, here is an excerpt from a student response: ``\emph{ML provided the quick answer when that was needed. Easier to use for single simulation than accessing supercomputer.}''

Students were asked to compare simulation-driven classroom teaching experience with a non-simulation-driven classroom teaching experience.
83\% of the students enjoyed simulation-driven teaching of computational science concepts stating that ``\emph{simulations aid to understanding the concept taught in the class more clearly}'' and ``\emph{it allows students to be the researchers and experimenters in the sense of using these tools to generate our results for our assignments}''.
The rest (17\%) still preferred the simulation-driven classroom teaching experience but stated that they felt there were occasions with extra downtime in the classroom because of waiting for cluster resources to run the simulations, and they suggested that ``\emph{the cluster waiting time needs to be filled with useful content}''.

Students were also asked to isolate what aspects of the nanoHUB online tool were most useful to them. 80\% of the responses indicated that the students like the ML prediction feature. Here is an excerpt from a student response: ``\emph{the predicted machine learning aspect was beneficial because it was very accurate with the simulated results, so if need be one do not have to wait for the simulation to finish computing to know what the results would have been}''. The survey responses also indicated that the students enjoy the freedom to probe the system behavior by tuning several model parameters, and they find the output graphs helpful. 
 
Finally, students were asked to provide suggestions to improve the ML-enhanced simulation tool. 66.6\% of the students provided feedback to improve the tool, while 33.4\% said that they do not have any suggestions to improve the tool. Some suggestions were: ``\emph{allowing users to download the ML prediction graph}'' and ``\emph{providing 3D snapshots of the simulation}''. The tool has been updated based on these useful suggestions and the latest version provides options to download the ML prediction result and visualize the snapshots of ions in confinement.
Some suggestions such as ``\emph{graph updates do not always happen when changing values and toggling ML, especially after full simulation was run}'' have not yet been implemented. These are related to the GUI rendering issues which we plan to resolve in the future working with the nanoHUB team.

\section{Discussion and Conclusion}

In this paper, we explored the potential of using ML surrogates for HPC-enabled simulations to address several educational challenges in teaching computational science and engineering courses. 
The ML surrogate yields predictions in excellent agreement with simulation, but at far less time and computing costs, delivering a dynamic and responsive simulation environment for rapid exploration by students in classroom settings. We developed a web application on nanoHUB that supported both HPC-enabled simulation and the ML surrogate methods to produce simulation outputs. 

The nanoHUB tool was used for both in-classroom instruction and for solving homework problems associated with two courses covering topics on computational materials science, modeling and simulation, and engineering applications of HPC-enabled simulations. 
The educational utility of the tool was evaluated using a survey that was answered enthusiastically by the students. Survey responses showed that the ML-enhanced tool is well-accepted among students and scored very high marks on convenience, user-friendliness, and consistency. Students also provided constructive feedback to improve the tool further in order to ensure its future success. The improvement in the interactivity with the simulation framework in terms of real-time engagement and anytime access enhanced the student learning of computational science concepts. The integrated simulation tool also enabled students to better understand the practical aspects of scientific computing including the tradeoffs between simulation accuracy, scalability, and efficiency.

Results from this investigation are encouraging and we expect the ML surrogate approach to be broadly applicable. We plan to explore the development of ML surrogates to predict outputs of other simulations including MD simulations of shape-changing nanoparticles \cite{brunk2019computational,brunk2020,video.shape} and different types of  Monte Carlo simulations \cite{frenkel,jadhao2010iterative1,jadhao2010iterative2}.
Another line of future work is to explore ways to reduce the training costs of the ML surrogates and probe their potential in predicting simulation outputs outside the pre-defined range of training datasets. 

We note that the integration of the ML surrogate in a computational tool hosted on nanoHUB exposes this approach to a much broader community of students, educators, and researchers. nanoHUB is the largest online resource for educational materials in nanotechnology \cite{klimeck.nanohub}, hosting over 500 web applications for launching simulations and serving over 1 million users worldwide. 

Finally, we want to emphasize that the use of ML surrogate is not intended to avoid or exclude HPC in education. Instead, our vision is for ML surrogate to complement and supplement HPC-enabled simulations for education applications. Note also that ML surrogate was designed using completed runs of HPC-enabled simulations. Without HPC, the time to generate the datasets to train the ML surrogate becomes prohibitively large \cite{kadupitige2019machine,kadupitiya2020machine2}. The use of ML surrogates contributes a novel way of teaching HPC topics by helping students develop intuition or ``feel'' for the physical system behavior through rapid exploration and visualization of variations in output quantities with changes in inputs, before they use HPC to solve specific problems.

\section*{ACKNOWLEDGMENT}
This work is supported by the National Science
Foundation through Awards 1720625 (Network for Computational Nanotechnology - Engineered nanoBIO Node) and DMR-1753182. Simulations were performed using the Big Red II supercomputing system. V.J. thanks G. C. Fox, F. Sun, and P. Sharma for useful conversations. We also thank all the students for providing valuable feedback on the ML-enhanced simulation tool.


\end{document}